\documentclass[aps,twocolumn,groupedaddress,superscriptaddress,amsmath,amssymb,prx]{revtex4-2}
\usepackage{amsmath}
\usepackage{graphicx}% Include figure files
\usepackage{dcolumn}% Align table columns on decimal point
\usepackage{bm}% bold math
\usepackage{bbold}
\usepackage[usenames,dvipsnames]{color}
\usepackage[colorlinks,citecolor=Blue,linkcolor=Red, urlcolor=Blue]{hyperref}
\usepackage{braket}
\usepackage[normalem]{ulem}

\DeclareUnicodeCharacter{2212}{\textendash}
\DeclareUnicodeCharacter{3B1}{\ensuremath{\alpha}}
\DeclareUnicodeCharacter{2009}{ }

\begin{document}
	
\author{A. Chiesa}
\affiliation{Dipartimento di Scienze Matematiche, Fisiche e Informatiche, Universit\`a di Parma, I-43124 Parma, Italy}    
\affiliation{UdR Parma, INSTM, I-43124 Parma, Italy}

\author{M. Chizzini}
\affiliation{Dipartimento di Scienze Matematiche, Fisiche e Informatiche, Universit\`a di Parma, I-43124 Parma, Italy}    

\author{E. Garlatti}
\affiliation{Dipartimento di Scienze Matematiche, Fisiche e Informatiche, Universit\`a di Parma, I-43124 Parma, Italy}
\affiliation{UdR Parma, INSTM, I-43124 Parma, Italy}

\author{E. Salvadori}
\affiliation{Dipartimento di Chimica \& NIS Centre, Universit\`a di Torino, Via P. Giuria 7, I-10125 Torino, Italy}

\author{F. Tacchino}
\affiliation{IBM Quantum, IBM Research -- Zurich, Säumerstrasse 4, 8803 Rüschlikon, Switzerland}

\author{P. Santini}
\affiliation{Dipartimento di Scienze Matematiche, Fisiche e Informatiche, Universit\`a di Parma, I-43124 Parma, Italy}
\affiliation{UdR Parma, INSTM, I-43124 Parma, Italy}

\author{I. Tavernelli}
\affiliation{IBM Quantum, IBM Research -- Zurich, Säumerstrasse 4, 8803 Rüschlikon, Switzerland}

\author{R. Bittl}
\affiliation{Freie Universität Berlin, Fachbereich Physik, Berlin Joint EPR Lab, Arnimallee 14, D‐14195 Berlin, Germany}

\author{M. Chiesa}
\affiliation{Dipartimento di Chimica \& NIS Centre, Universit\`a di Torino, Via P. Giuria 7, I-10125 Torino, Italy}

\author{R. Sessoli}
\affiliation{Dipartimento di Chimica “Ugo Schiff” \& INSTM, Universit\`a Degli Studi di Firenze, I-50019 Sesto Fiorentino}

\author{S. Carretta}
\email{stefano.carretta@unipr.it}
\affiliation{Dipartimento di Scienze Matematiche, Fisiche e Informatiche, Universit\`a di Parma, I-43124 Parma, Italy}
\affiliation{UdR Parma, INSTM, I-43124 Parma, Italy}

%\title{Chiral-induced spin-selectivity in electron-transfer processes: 	fingerprints from magnetic resonance}

\title{Assessing the nature of chiral-induced spin-selectivity by magnetic resonance}


\begin{abstract}
Understanding chiral induced spin-selectivity (CISS), resulting from charge transport through helical systems, has recently
inspired many experimental and theoretical efforts, but is still object of intense debate. 
In order to assess the nature of CISS, we propose to 
focus on electron-transfer processes occurring at the single-molecule level. We design simple magnetic resonance experiments, exploiting a qubit as a highly sensitive and coherent magnetic sensor, to provide clear signatures of the acceptor polarization. 
Moreover, we show that information could even be obtained from time-resolved electron paramagnetic resonance experiments on a randomly-oriented solution of molecules.
The proposed experiments will unveil the role of chiral linkers in electron-transfer and could also be exploited for quantum computing applications.

\end{abstract}                              
\maketitle

%\twocolumngrid 

\indent 

%\section{Introduction}

\begin{figure}[h]
\centering
\includegraphics[width=1\linewidth]{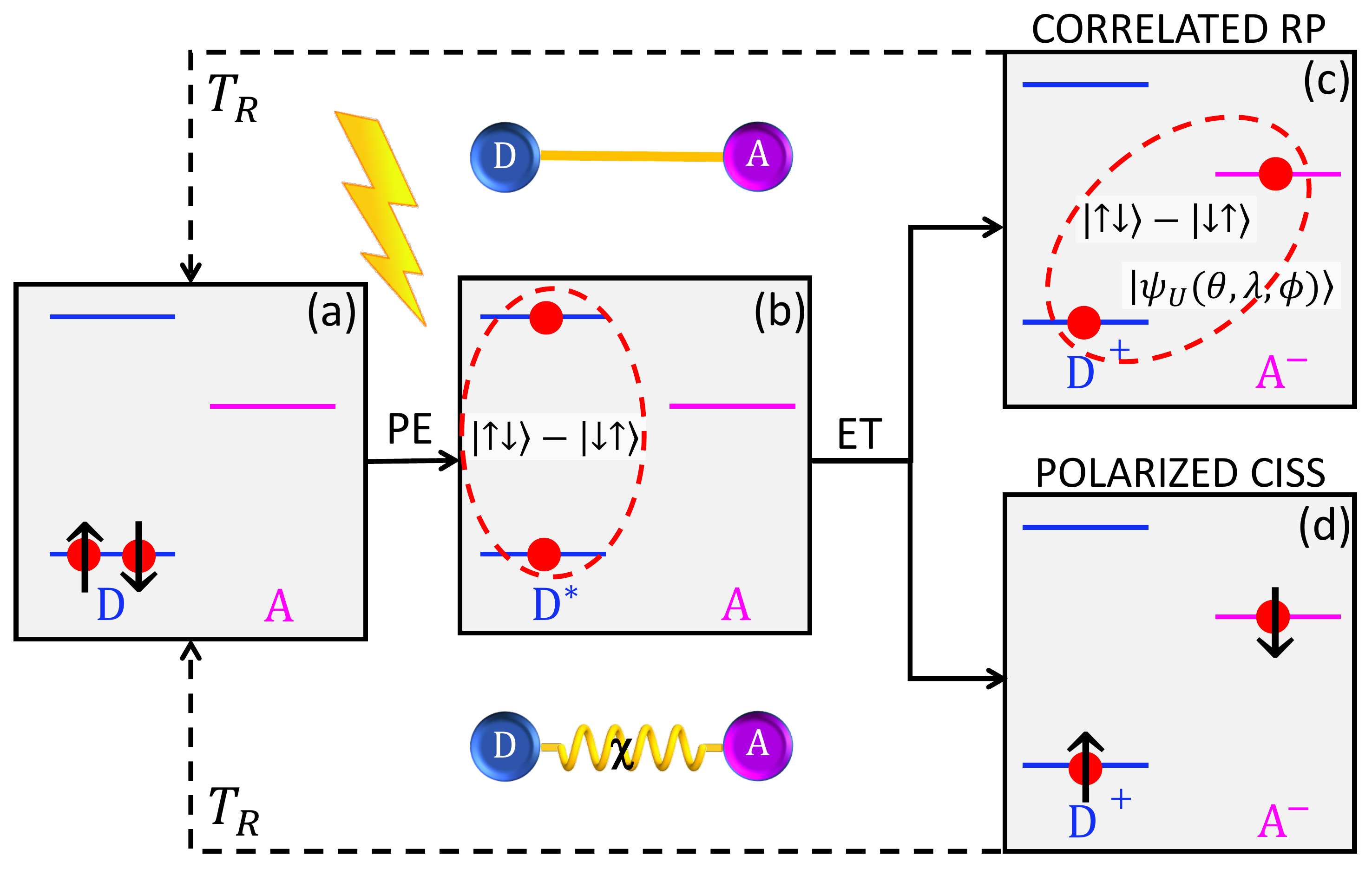}
\caption{Scheme of the electron-transfer mechanism: (a) Singlet initial state on the donor D (with two electrons both in the ground orbital). Photo-excitation (PE) brings to the D$^*$A singlet state in which one electron is excited (b), but the pair is still in an entangled state (dashed circle). After electron-transfer (ET) of the excited electron to the acceptor (A) the final state is still either a correlated radical pair (RP) or a polarized state after transfer through a chiral bridge. Recombination to the initial singlet (or to the triplet) state occurs on a time scale $T_R$ (dashed arrows). Top (bottom) inset: Scheme of the DA radical pair, linked by a linear (chiral) bridge.} 
\label{Fig1et}
\end{figure}

Charge displacement through chiral systems has been suggested as a resource for spintronics devices and as the driving force of many biological reactions \cite{RevNaaman19,NaamanAcc20,ACIEN20}. This has led to huge research efforts, mainly focused on transport setups in which chiral molecules are 
deposited on a FM electrode\cite{annurev-physchem15} and the spin polarized current is filtered by the chiral molecules, a phenomenon known as chiral-induced spin selectivity (CISS). 
Most experiments were done on 
self-assembled monolayers of chiral molecules ($\chi-$SAM), 
but individual molecules were also addressed by atomic force microscopy  \cite{Nogues2004,Xie2011,Small17,Waldeck2021}.
Several studies, concerning a $\chi-$SAM on a conducting substrate, revealed polarization also in very different contexts, in which chiral molecules are in a static, even if out-of-equilibrium, configuration \cite{NcommNaaman17,Banerjee-Ghosh1331,Kumar17,Smolinsky}.
In parallel, different theoretical models have been put forward \cite{Guo2014,Varela2016,Pan2016,Balseiro2016,MujicaMaster,Cuniberti2019,Herrmann2020,Varela2020,Medina2015,NaamanTh,Cuniberti2020,Fransson2019,Dalum2019,Polaron,Peijia2020,Liu2021}, 
but a comprehensive, even qualitative, description of the phenomenon is still lacking \cite{Waldeck2021}. \\
%In particular, although the above observations indicate CISS as a property of individual chiral molecules, it is still object of intense debate weather the phenomenon emerges only for unbound initial states \cite{Dalum2019,Fay2021} (as in $\chi$ molecules embedded between FM leads) or also in molecular states. \\
To shed light into the origin of CISS, we need to simplify as much as possible the experimental setup and focus on qualitative features, emerging directly from chiral molecules. In particular, electron-transfer (ET) processes through a chiral bridge linking a donor and an acceptor (D-$\chi$-A in the following, see bottom inset of Fig. \ref{Fig1et}) may serve as the ideal platform to understand this phenomenon. 
Recently, a minimal model of ET in chiral environments was proposed, in which the bridge was included through an effective spin-orbit interaction, inducing a coherent rotation of the transferred electron spin \cite{Fay2021}. Starting from the singlet state precursor obtained by photo-excitation (PE), this model does not predict any local polarization on D/A.
In contrast, experiments on photosystem-I \cite{RevNaaman19,Carmeli14} have demonstrated a spin polarization occurring also in ET processes. \\
We propose here simple experiments to unambiguously distinguish the two situations, thus finally elucidating the nature of CISS, by answering the question: is the electron polarized after ET through the chiral bridge? 
The experiments are based on using a highly coherent qubit (Q), coupled to the acceptor in a D-$\chi$-A-Q setup, as a local probe of this polarization transfer and time-resolved %pulse
electron paramagnetic resonance (TR-EPR) as the experimental tool. By acting as an external and local sensor, the qubit gives direct access to the acceptor polarization, without influencing the ET process. This provides a unique means to assess the nature of CISS at the single molecule level, much more directly compared to previous setups \cite{RevNaaman19}, where many additional ingredients could somehow darken the role of the $\chi$ unit. 
%Similar information can be achieved by using a nuclear spin $I$ (coupled to the acceptor by hyperfine interaction) as an alternative probe, in a nuclear magnetic resonance (NMR) apparatus. 
The second experiment we propose probes the qubit state after polarization has been coherently transferred from A to Q by an appropriate pulse sequence.
%We show below that both approaches implemented on an oriented solution of D-$\chi$-A-Q molecules yield unambiguous fingerprints of the polarization of the acceptor \comm{and that clear hints of such polarization also emerge from TR-EPR on a frozen solution.} 
We show that both approaches yield unambiguous fingerprints of the polarization of the acceptor if implemented on an oriented solution of D-$\chi$-A-Q molecules. % \comm{and that clear hints can even be obtained from experiments on a frozen solution}.
Moreover, we demonstrate by numerical simulations that quantitative %(although not qualitative) 
features of CISS can also be observed in TR-EPR spectra of a much simpler experimental setup consisting of a randomly oriented ensemble of D-$\chi$-A molecules. 
%this procedure can be harnessed in quantum technologies, enabling qubit initialization and read-out and hence room-temperature quantum computation, as demonstrated by extensive numerical simulations.

{\it D-$\chi$-A system --}
We consider the following experimental scenario, schematically shown in Fig. \ref{Fig1et}: the donor, initially in a doubly-occupied ground state (panel a) is photo-excited to the D$^*$-$\chi$-A singlet state (b).
Then, ET yields the charge-separated state D$^+$-$\chi$-A$^-$. In the case of a linear bridge linking donor and acceptor (top inset), this is expected to be still a two-electron singlet state (Fig. \ref{Fig1et}-(c)). Our aim is to compare this situation, typical of a spin-correlated radical pair (RP) \cite{Stehlik1989,Salikhov1990,Bittl1991,Kothe1991,Kothe1994,Weber1995,Weber2010}, with that in presence of a chiral bridge.
In particular, in the experiments proposed below we consider the following charge separated states:
\begin{enumerate}
    \item A singlet state, $|\mathcal{S}\rangle = (|{\uparrow\downarrow}\rangle - |{\downarrow\uparrow}\rangle)/\sqrt{2}$, typical of ET through a linear bridge.
    \item A polarized state, 
    represented by the density matrix $\rho_p = \frac{1+p}{2}  | {\uparrow\downarrow}\rangle \langle {\uparrow\downarrow}| + \frac{1-p}{2} |{\downarrow\uparrow}\rangle \langle {\downarrow \uparrow}|$ (with  $-1 \le p \le 1$ and $\neq 0$). 
    Here $p = -2{\rm Tr}[\rho_p S_{zA}]$ is the final polarization of the acceptor. %, with the sign depending on $\chi$ handedness.
    This state (represented in Fig. \ref{Fig1et}-(d) for $p=1$) could result from spin selectivity after ET through a chiral bridge, as found in measurements on photosystem-I \cite{Carmeli14} and in other experiments on $\chi-$SAM 
    \cite{NcommNaaman17,Banerjee-Ghosh1331,Ghosh2020,Kumar17,Smolinsky}, where charge polarization was induced by application of an electric field, thus making these situations somehow similar to ET. 
    \item A non-polarized (correlated) state $|\psi_U\rangle$ (Fig. \ref{Fig1et}-(c)) resulting from a coherent rotation of the transferred electron belonging to $|\mathcal{S}\rangle$, as proposed in \cite{Fay2021}. The most general form of this state is given by  $|\psi_U \rangle = {\rm cos} \frac{\theta}{2} \frac{e^{i(\lambda+\phi)}|\uparrow\downarrow\rangle -  |\downarrow\uparrow\rangle}{\sqrt{2}}-{\rm sin} \frac{\theta}{2}
    \frac{e^{i\lambda}|\uparrow\uparrow\rangle + e^{i \phi} |\downarrow\downarrow\rangle}{\sqrt{2}}$.
    One can easily check that $|\psi_U \rangle$ does not give any local polarization, i.e. $\langle \psi_U | S_{zD} | \psi_U \rangle = \langle \psi_U | S_{zA}|\psi_U \rangle = 0$.
\end{enumerate}
%\comm{PEZZO MODEL ERA QUI.}
%To understand the origin of the these different states...
The three possible ET outputs are clearly discriminated by the experiments proposed below.

{\it Detecting polarization using a qubit sensor --}
\begin{figure*}[ht]
\centering
\includegraphics[width=0.9\linewidth]{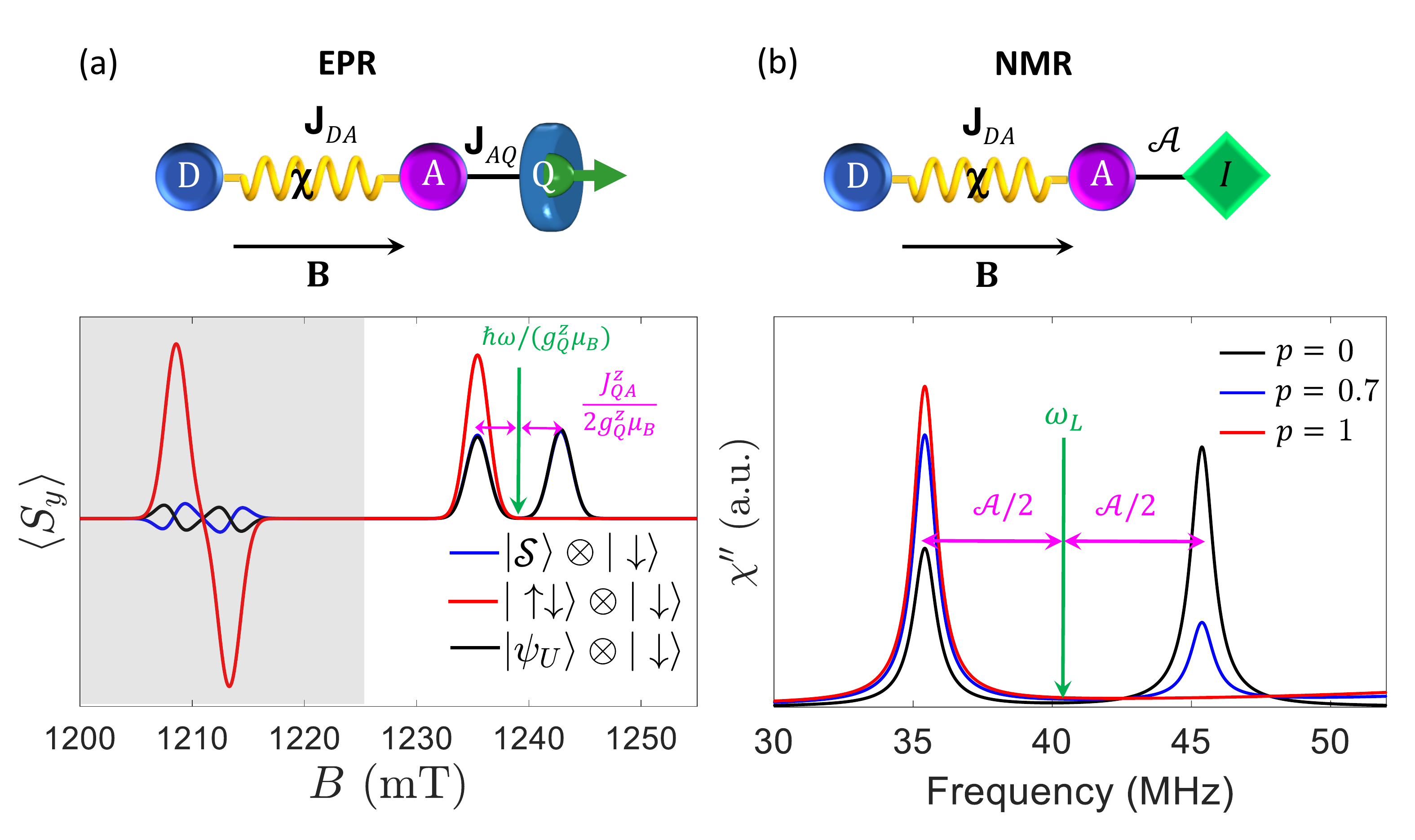}
\caption{(a) Q-band TR-EPR spectrum of a D$^+$-$\chi$-A$^-$-Q system (sketched on top) with the chiral axis aligned parallel to the external field (integrated from 100 to 300 ns) and for different initial states of the radical pair (with the qubit always in $|\downarrow\rangle$): singlet (blue line), corresponding to transfer along a linear bridge without CISS effect; fully polarized state (on both donor and acceptor, red); un-polarized state $|\psi_U\rangle$ as suggested in Ref. \cite{Fay2021}. The grey-shaded area represents the signal from donor-acceptor, while at larger field the absorption peaks are due to the qubit. 
(b) NMR spectrum as a function of frequency, probing nuclear excitations on a nuclear spin 1/2 (e.g. a $^{19}$F, Larmor frequency $\nu_L \approx 40$ MHz at 1 T) coupled by hyperfine interaction to the donor. The different intensity of the two peaks for $p=0$ is due to different matrix elements for the two transitions. Variance from the $p=0$ behavior directly measures the acceptor polarization.
Parameters: $h \nu = 34$ GHz, $J_z \approx 200$ MHz, $r_{DA} = 25 ~\AA$, $r_{AQ} = 8 ~\AA$, $A=10$ MHz, $g_{1,2} = g_e \mp \Delta g/2$, with $\Delta g = 0.002$, ${\bf g}_Q = (1.98, 1.98, 1.96)$, as typical for VO$^{2+}$ or Ti$^{3+}$ \cite{Camargo2021}. 
Inhomogeneous broadening of the parameters is included by a gaussian broadening of the peaks with FWHM 2.35 mT. To generalize our analysis, we did not include parameters of a specific qubit, such as hyperfine interaction. }
\label{Fig2}
\end{figure*}
To detect spin unbalance in the D$^+$-$\chi$-A$^-$ unit, we propose the use of a qubit, i.e. a paramagnetic $S=1/2$ center showing long coherence. 
%Qubits have already been proposed as very sensitive magnetometers. 
Molecular spin qubits are very promising sensors\cite{RMPSensing} thanks to the very long coherence times they can reach if properly chemically engineered \cite{Zadrozny,Freedman_JACS,Freedman2014,Freedman_Ni,Freedman_Cr,Atzori_JACS,Atzori2016,SIMqubit,Bader,Hill,Atzori2017,Atzori2018,VOTPP,Sessoli2019}. In addition, the capability to link these qubits to other units make them ideal candidates for the proposed architecture.
As a model example, we consider a VO qubit, which has already demonstrated remarkable coherence even at room temperature \cite{Atzori2016}
also with porphyrin-based ligands \cite{VOTPP,Urtizberea} that con be easily functionalized to couple with the D-$\chi$-A dyad \cite{Atzori_JACS}.
Interestingly, a setup consisting of a chain of three radicals was studied in Refs.  \cite{Kandrashkin1997,Salikhov1999,Salikhov2002,Kandrashkin2007,Wasielewski2013,rugg2019}. \\ 
%Here the role of the qubit is substantially different, acting as coherent quantum sensor which does not perturb the RP but only detects local spin polarization.} \\
After ET, the whole D-$\chi$-A-Q system is described by the following spin Hamiltonian:
\begin{equation}
        H_{S} = \mu_B \sum_{i=D,A,Q}  {\bf S}_i \cdot {\bf g}_i \cdot {\bf B} + \sum_{i,j} {\bf S}_i \cdot {\bf J}_{ij} \cdot {\bf S}_j,
\label{Hams}
\end{equation}
where the first term models the interaction of each of the three spins with an external magnetic field, while the second describes the magnetic dipole-dipole interaction between D-A and A-Q. We assume the point-dipole approximation, leading to ${\bf J}_{ij} = [ {\bf g}_i \cdot {\bf g}_j -3 ({\bf g}_i \cdot {\bf r}_{ij})({\bf g}_j \cdot {\bf r}_{ij})]\mu_B^2/r_{ij}^3$, and in the following we consider for simplicity a linear 3-spin chain (see Fig.\ref{Fig2}-(a)), with $r_{DA} = 25 ~$\AA, and isotropic $g_{D,A}$. Possible additional isotropic contributions to ${\bf J}_{ij}$ do not alter our conclusions, provided that $J_{AQ}^{x,y}$ is sufficiently smaller than $|g_A-g_Q^z|\mu_B B$ to make the initial state of the qubit factorized from that of the acceptor. In this regime, Q acts as a coherent quantum sensor which does not perturb the RP but only detects local spin polarization. At the same time, $J_{AQ}^z$ should be larger than the linewidth corresponding to Q excitations. 
For instance, by choosing FWHM = 2.35 mT (a conservative estimate for typical transition metal ion based qubits \cite{Atzori_JACS,Camargo2021}), these conditions are easily fulfilled with 6~\AA $\lesssim r_{AQ} \lesssim 11 ~$\AA~ and working in Q-band. We thus fix $r_{AQ} = 8~$\AA~ in the following simulations.\\
% and then the system evolves subject to Hamiltonian (\ref{Hams}) plus additional incoherent effects \cite{Bittl1991,Kothe1994} (see below). \\
Two different experiments exploiting the qubit as a sensor of the acceptor polarization are proposed. The first consists of TR-EPR measurements recorded immediately after ET. 
To simulate TR-EPR spectra, we compute the time evolution of the system density matrix by integrating the Liouville equation 
$\dot{\rho} = -i\left[ \frac{1}{\hbar}\tilde{H} +i \tilde{R} +i \tilde{K} \right] \rho$, where $\rho$ is the system density matrix in the rotating frame and 
$\tilde{H}$, $\tilde{R}$ and ${\tilde{K}}$ are super-operators associated to the system Hamiltonian (including also a continuous-wave oscillating field) and to phenomenological relaxation/decoherence and recombination mechanisms, respectively (see SI and e.g. \cite{Kothe1991,Bittl1991,ZWANENBURG1993,Kothe1994,Weber1995,Weber2010}).
In particular, we assume the same relaxation ($1/T_1$) and dephasing rates ($1/T_2$) for each
of the three spins (treated as independent as in Ref. \cite{Salikhov2002}), 
%diagonal element and the same dephasing rate ($1/T_2$) for each off-diagonal element of the RP density matrix, in the singlet-triplet RP basis, 
using conservative values (even at room temperature) of $T_1 \sim 2 \, \mu$s, $T_2 \sim 0.5\,\mu$s \cite{Kothe1991,Kothe1994,Atzori2016}. Finally, the recombination rate $T_R$ for the radical pair is assumed in the $10 \, \mu$s range. 
%super-operator is given by $K_{ijkl} = -1/T_R (\delta_{ik} \delta_{jl} + \delta_{ij} \delta_{ik})$, with $1/T_R$ being the charge recombination rate, assumed in the $10 \, \mu$s range. 
We stress that all simulations have been performed using realistic parameters and neglecting inter-system crossing, a safe choice for these systems characterized by fast ET.
The recorded signal then corresponds to $\langle S_y(t,B) \rangle = {\rm Tr} [\sum_i S_{yi} \rho(t)]$.
In order to unambiguously unveil the nature of CISS, we consider an oriented solution of D-$\chi$-A-Q molecules, with the static field parallel to the chiral axis (experimentally achieved for instance by poling, thanks to the large electric dipole moment typical of chiral molecules based on oligopeptides \cite{DipoleDNA,NaamanTh,Wallace2006,Varela2019}). \\
%\comm{However, starting from a well characterized qubit with separated excitations along different directions, signatures of A polarization can be obtained also by working on a frozen solution of  D-$\chi$-A-Q molecules, as shown in the SI.}  \\
The three states (1-3) give distinct TR-EPR signals, as shown in the time-integrated spectra of Fig. \ref{Fig2}-(a). In particular, we note that different (un-polarized) states, such as $|\mathcal{S}\rangle$ or $|\psi_U (\theta ,\phi,\lambda)\rangle$, modify the radical-pair spectrum (black vs. blue curves in the grey-shaded area), but not the qubit response (right part of the spectrum), which is only affected by $\langle S_{zA} \rangle$. The qubit absorption peak close to $\sim 1.24$ T is split by the interaction with the acceptor. If the latter is completely polarized (e.g. in $|{\downarrow}\rangle_A$ state, $p=1$), a single peak appears, corresponding to the $|{\uparrow\downarrow\downarrow}\rangle \rightarrow |{\uparrow\downarrow\uparrow}\rangle$ transition \footnote{We have assumed here a completely polarized initial state of the qubit, as could be obtained at Q-band at low temperature, but partial population of the excited qubit state only yields an overall reduction of the qubit signal.}. Conversely, an un-polarized DA state induces the additional excitation corresponding to  $|{\uparrow}\rangle_A$ with approximately the same intensity (apart from slightly different matrix elements or thermal population). 
Note that in the present simulation, $\langle S_y(t,B)\rangle$ shows a weak time dependence (see SI), thus making the choice of the time window of integration not crucial. \\
\begin{figure}[t]
\centering
\includegraphics[width=0.95\linewidth]{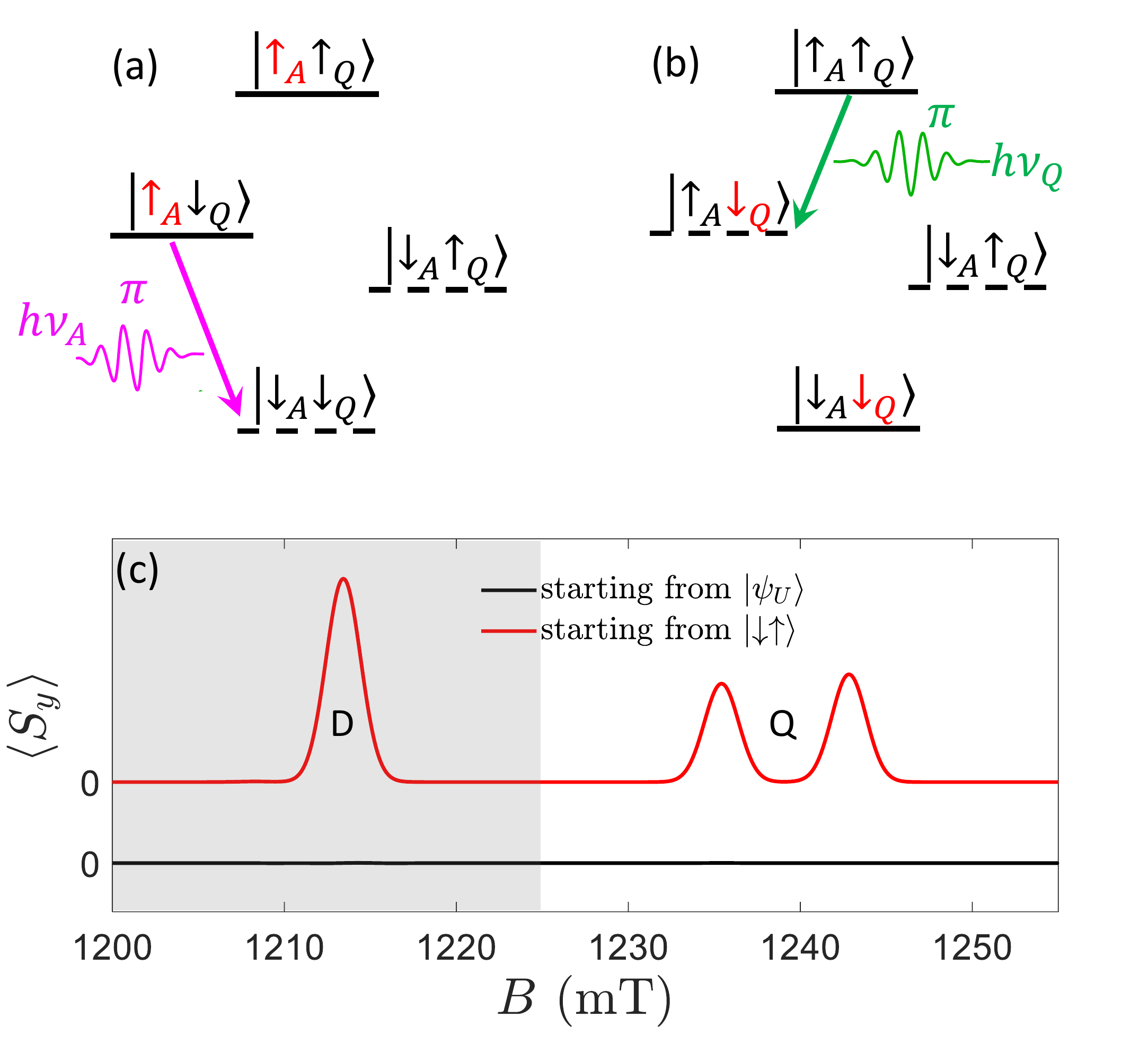}
\caption{Polarization transfer to the qubit probe: (a,b) Pulse sequence implementing the scheme on the AQ pair (D has opposite polarization compared to A and is not affected by the pulses, i.e. rotations of A are independent from state of D). Full (dashed) lines indicate occupied (empty) states, initially with fully polarized A, finally with polarization transferred to Q (red arrow). The two pulses on A (Q) are indicated by a purple (green) arrow.
(c) TR-EPR spectrum (integrated from 100 to 300 ns) after application of the polarization transfer sequence for a un-polarized state (black line) or for a spin polarized one (red), as expected after CISS. Transitions involving excitations of D (Q) are represented by peaks at low (high) field.  }
\label{Fig3}
\end{figure}
A similar information can be obtained by broadband NMR spectroscopy, as reported in Fig. \ref{Fig2}-(b). We consider, in this case, a nuclear spin $I=1/2$ (such as $^{19}$F) coupled to A by (isotropic) hyperfine interaction $\mathcal{A} ~ {\bf S}_A \cdot {\bf I}$. The NMR absorption signal ($\chi^{''}$) is shown as a function of frequency, in the range corresponding to the excitation of nucleus $I$, split by hyperfine interaction with A. Again, the probe is only sensitive to $\langle S_{zA} \rangle$ and weakly perturbs the system, thus giving direct access to the acceptor polarization \footnote{Detection of NMR spectrum must be sufficiently faster compared to relaxation/recombination times. Careful choice of the nucleus used as a probe and/or deuteration could be necessary to reduce broadening of the probe peaks due to the interaction with other spins (which must be smaller than $\mathcal{A}$) and to isolate the probe Larmor frequency.}.\\
%\comm{Add comment on ENDOR? Detection could be faster than with NMR. We need to excite A to probe the linked nucleus or to consider Q hyperfine coupled to a nucleus? Less direct measurement...}\\
Partial polarization leads to intermediate situations (blue curve in Fig. \ref{Fig2}-(b)), thus making the relative intensity of the two peaks a measure of spin polarization. 
Remarkably, both for EPR and NMR experiments, this feature is not hampered by performing the experiment at high temperature, which only induces an overall attenuation of the signal (see simulation in the SI). 
Moreover, opposite polarization (arising in model (2) by changing the enantiomer) yields inversion of the intensity of the two peaks, thus providing a direct proof of the occurrence of CISS.\\
\begin{figure}[t]
\centering
\includegraphics[width=1\linewidth]{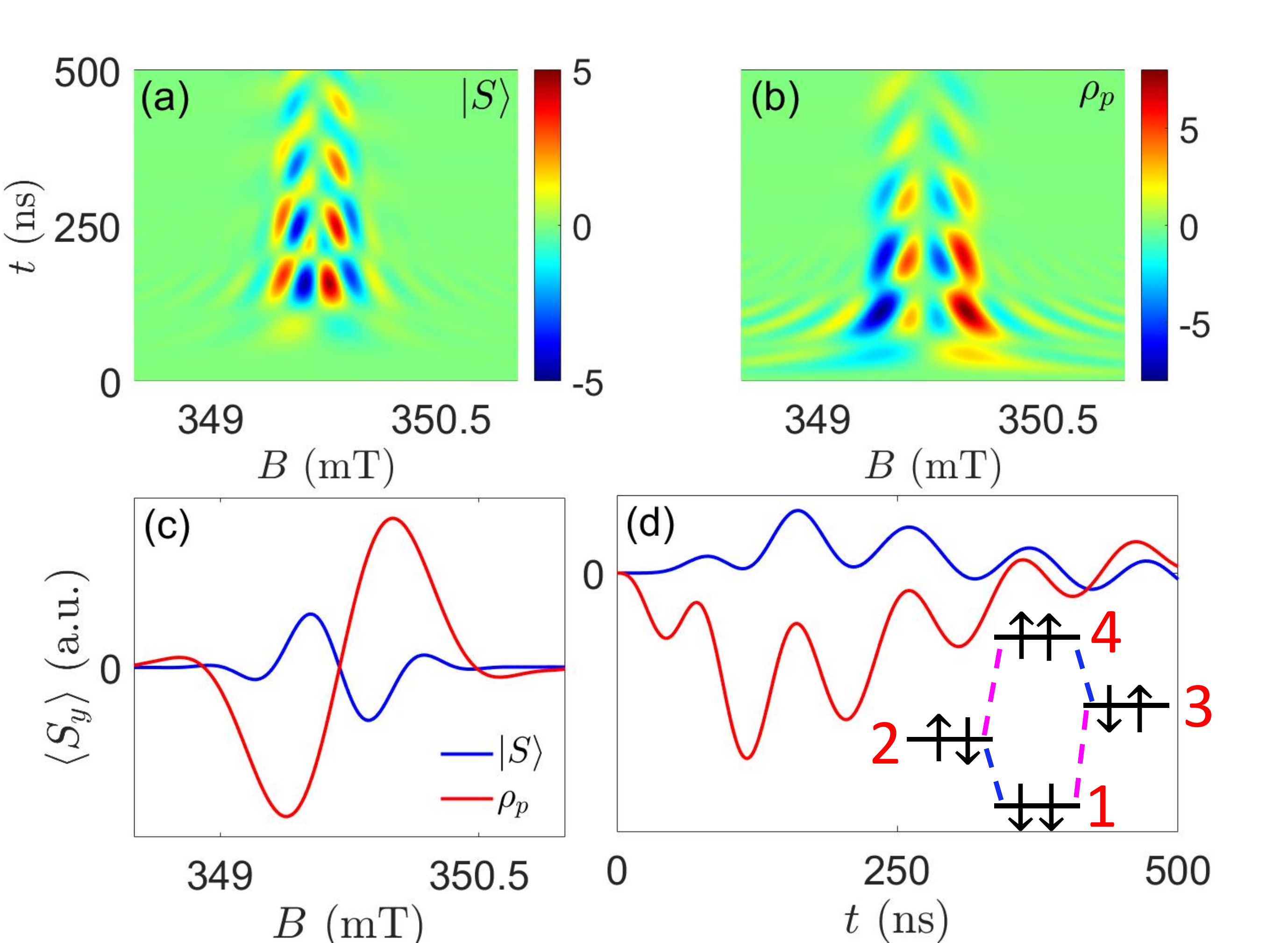}
\caption{TR-EPR on a randomly oriented ensemble of D-$\chi$-A molecules. Parameters: $\Delta g = 0.002 $, $r_{DA} = 25~ \AA$, $h \nu = 9.8$ GHz. (a,b) Two-dimensional maps of $\langle S_y (t,B) \rangle$ for an initial singlet or polarized state, respectively. (c) Field dependence of the absorption TR-EPR spectrum, integrated in the time-window corresponding to the first maxima-minima in the maps of panels (a,b), for the states $|\mathcal{S}\rangle$, $\rho_p$ (with either $p=-1$ or $p=1$, leading to the same result in solution).  (d) Time dependence around $B \approx 349.5$ mT, highlighting the opposite behavior at short times for polarized and un-polarized states. Simulations include relaxation, dephasing and recombination of the radical pair, with $T_1=2 ~\mu$s, $T_2=0.5 ~\mu$s, $T_R=10 ~\mu$s (in the singlet-triplet RP basis, see SI), and gaussian broadening with FWHM = 0.15 mT \cite{Bittl1991}. Inset: schematic energy-level diagram, with states practically corresponding to eigenstates of $S_{zi}$. Allowed EPR transitions of D (A) are indicated by blue (purple) dashed lines.}
\label{Fig4}
\end{figure}
The second experiment is based on 
a sequence of pulses properly designed to coherently transfer polarization from A to Q, followed by EPR measurement of the final state of the system.
The sequence is reported in Fig. \ref{Fig3}-(a,b): it consists of a first $\pi$-pulse on A (conditioned by Q state, purple arrow), followed by a $\pi$-pulse on Q, conditioned by the state of A (green arrow). Starting with the qubit in a complete mixture, $\rho_Q = \mathbb{I}/2$ (high-temperature limit), and a fully polarized acceptor state, this sequence completely transfers polarization to Q, leaving A in an un-polarized mixture. 
The final state of the system can be measured by EPR after the pulses. The resulting $\langle S_y \rangle$ is reported in Fig. \ref{Fig3}-(c), for the case of an initially polarized acceptor state (red line), compared to an initially un-polarized one ($|\psi_U\rangle$ or $|\mathcal{S}\rangle$, black). The latter gives a very weak signal, both at low field (from the radical pair) and at high field (from the qubit). Conversely, starting from a polarized $|{\downarrow_D \uparrow_A}\rangle$ state, the final state is also polarized on both D and Q. The peaks corresponding to excitations of Q and D are then split by ${\bf J}_{AQ}$ and ${\bf J}_{DA}$, respectively (the latter being much smaller and hence not visible in Fig. \ref{Fig3}-(c)).  %We finally note that no significant difference between an initial $|S\rangle$ or $|\psi_U\rangle$ state is found in the TR-EPR spectrum. \\
Hence, a qubit (or a nuclear spin) weakly interacting with a D-$\chi$-A unit is the ideal probe of the spin imbalance on the DA pair and would unveil the nature of CISS. We also point out that 
the proposed platform is robust even at room temperature, a condition in which many $\chi$ units keep
large polarization efficiency \cite{NaamanAcc20} and VO qubits maintain remarkable coherence \cite{Atzori2016}. 

%\comm{Inter-system crossing potrebbe essere un problema?}\\
{\it TR-EPR on D-$\chi$-A in solution --}
In order to facilitate the first experimental attempts, we further simplify our setup and consider
TR-EPR experiments \cite{Bittl1991, Kothe1991,Kothe1994,Mattias,Kandrashkin2007,Wasielewski2013} 
on an isotropic solution of D-$\chi$-A molecules.
It was recently pointed out \cite{Hore2021} that angular average on the initial state cancels the most clear signature of CISS. %, leaving only a scale factor difference between a polarized and a singlet precursor. However, if both the initial state and the Hamiltonian break spherical symmetry, simulations must include an average on the system density matrix after time evolution. 
However, we find that, in presence of an anisotropic dipolar DA interaction, quantitative signatures of CISS are already present in the spectrum detected on an isotropic solution of D-$\chi$-A molecules. 
%These features could be even enhanced by anisotropy in ${\bf g}_i$ (see SI).}  
In particular, the different initial states discussed in the previous sections lead to different time evolutions and hence to significantly different spectra at short times. 
%For longer times, the oscillation frequency of $\langle S_y(t) \rangle$ depends on the Hamiltonian parameters and hence characterization of the system becomes necessary to understand the long-time dynamics.
As an example, Fig. \ref{Fig4} shows simulated TR-EPR spectra at 9.8 GHz (X-band), along with cuts for specific time/field windows.
We immediately note that at short times the $|\mathcal{S}\rangle$ state (panel a) gives an opposite pattern of maxima and minima, compared $\rho_p$ (panel b). 
This also emerges from the short-times spectra, as a function of $B$ (panel c), and from the time dependence, reported in panel (d) for $B$ corresponding to the first pronounced peak. 
The different order of maxima and minima as a function of $B$ can be understood by considering the form of the initial state along different directions. 
An illustrative diagram of the (practically factorized) eigenstates is shown in the inset of Fig. \ref{Fig4}-(d), with levels labeled in order of increasing energy from 1 to 4 and allowed transitions indicated by dashed lines.
%of D (A) (purple) arrow. 
Note that the corresponding gaps and resonance fields are made different by {\bf J}$_{DA}$.
An initial $|\mathcal{S}\rangle$ state shows spherical symmetry and hence keeps the same form in any direction. The static Hamiltonian induces partial population transfer from the dark $|\mathcal{S}\rangle$ state to the symmetric superposition $|T_0\rangle = (|{\uparrow\downarrow}\rangle+|{\downarrow\uparrow}\rangle)/\sqrt{2}$. Hence, the EPR signal shows emission lines for 2-1 and 3-1 transitions and absorptions for 2-4 and 3-4.
Conversely, an initial $\rho_p$ state (polarized along the chiral axis) is strongly anisotropic. For a generic orientation of the molecule with respect to the external field (which defines the quantization axis) such a state has sizable components on states $|{\uparrow\uparrow}\rangle$ and $|{\downarrow\downarrow}\rangle$. If these components are larger than $|T_0\rangle$ we get emission lines for 4-2 and 4-3 transitions and absorption for 1-2 and 1-3. This situation (opposite to that of the singlet) dominates on the spherical average (see SI for details). 
Then, the order of maxima and minima is determined by the sign of the spin-spin interaction (fixed by considering a dipolar coupling).\\
%The situation is similar starting from a $|{\uparrow\downarrow}\rangle$ state along the external field ($z$) axis. Conversely, an initial state polarized along $x,y$ has only components on states $|{\uparrow\uparrow}\rangle$, $|{\downarrow\downarrow}\rangle$ and $|\mathcal{S}\rangle$ \footnote{This states can be written in the $z$ basis by computing $e^{-i S_{y,x}\pi/2} |{\uparrow\downarrow}\rangle$.}. This gives emission lines for 4-2 and 4-3 transitions and absorption for 1-2 and 1-3, a situation opposite to the singlet-born spectrum on two out of three directions and hence also on the spherical average. Then, the order of maxima and minima is determined by the sign of the spin-spin interaction (fixed by considering a dipolar coupling).}\\
We finally note that, since the spectrum in panel (b) is the same for any choice of $p$, this measurement does not probe the acceptor polarization, but distinguishes a factorized state $\rho_p$ from a singlet.
As shown in the SI, 
un-polarized states $|\psi_U\rangle$ give different patterns, intermediate between $|\mathcal{S}\rangle$ and $\rho_p$, thus requiring, in general, a preliminary characterization of the system Hamiltonian to clearly reveal CISS.  \\
In conclusion, we have proposed simple magnetic resonance experiments exploiting a qubit as a probe of the acceptor polarization in electron-transfer processes through a chiral bridge. These experiments will ultimately unravel the nature of chiral-induced spin selectivity at the single-molecule level. 
We finally note that, by applying the proposed sequence for polarization transfer, the CISS effect could be exploited for initialization and read-out of the qubit state, 
alternative to optical initialization recently achieved in a Cr$^{(IV)}$ $S=1$ complex \cite{Bayliss1309}. This is 
a crucial step towards the physical implementation of quantum computers. \\

%To understand the origin of the states (1-3) proposed above, we need to explicitly include the chiral bridge in our description. 

%\acknowledgements{
This work received financial support from the Italian Ministry of Education and Research (MUR) through PRIN project 2017 Q-chiSS “Quantum detection of chiral-induced spin-selectivity at the molecular level” 
and from the European Union's Horizon 2020 program under Grant Agreement No. 862893 (FET-OPEN project FATMOLS).%}

%\section{Author contributions}

%\section{Competing interests}
%The authors declare no competing interests.\\

\end{document}